\documentstyle[12pt, fleqn, psfig]{article}
\newcommand{\be}{\begin{equation}}
\newcommand{\ee}{\end{equation}}
\newcommand{\ba}{\begin{eqnarray}}
\newcommand{\ea}{\end{eqnarray}}
\newcommand{\pa}{\partial}
\newcommand{\Om}{\Omega}
\newcommand{\ds}{\displaystyle}
\newcommand{\bm}{\boldmath}

\makeatletter                                               
\def\@eqnnum{{\normalfont \normalcolor [\theequation]}}
\makeatother

\makeatletter
\def\@cite#1#2{({#1\if@tempswa , #2\fi})}
\makeatother

\makeatletter
\def\@biblabel#1{#1.}
\makeatother

\begin{document}
\title{Interaction of two deformable viscous drops under external temperature gradient} 
\author{ V. Berejnov, O. M. Lavrenteva, and A. Nir \\
{\small {Department of Chemical Engineering, Technion - Israel Institute of
Technology,}}\\
{\small {Haifa 32000, Israel}}}
\maketitle






\begin{abstract}{The axisymmetric deformation and motion of interacting droplets in
an imposed  temperature gradient is considered using boundary-integral techniques
for slow viscous motion. Results showing temporal drop motion, deformations and
separation are presented for equal-viscosity fluids. The focus is on cases when
the drops are of equal radii or when the smaller drop trails behind the larger
drop. For equal-size drops, our analysis shows that  the motion of a leading drop
is retarded while the motion of the trailing one is enchanced compared to the
undeformable case. The distance between the centers of equal-sized
deformable drops decreases with time. When a small drop follows a large one, two 
patterns of behavior may exist. For moderate or large initial separation the
drops separate. However, if the initial separation is small there is a transient 
period in which the separation distance initially decreases  and only afterwards 
the drops separate. This behavior stems from the multiple time scales that exist
in the system.
\bigskip
\bigskip
\bigskip

}
\end{abstract}

\newpage
\bigskip

\section{Introduction}

\medskip
It has long been known that a viscous droplet submerged into another immiscible
non-isothermal liquid migrates in the direction of the temperature  gradient,
$\nabla T$. This droplet's migration is due to the temperature-induced surface
tension gradient ${\nabla_s} {\Large (\gamma)}$ at the droplet interface $S$ and it
is addressed as a thermocapillary migration.  Young {\it et al}
\cite{YGB59} were first to investigate the  mechanism of such migration 
both theoretically and experimentally. These authors found 
conditions for the stabilization of a buoyant
bubble under gravity force by the thermocapillary force which was directed
oppositely. They obtained a theoretical prediction for the migration velocity of a
single spherical bubble of radius $a$, which is placed in a viscous fluid of viscosity
$\eta_1$, with an imposed constant temperature gradient ${\bf A}$. In the limit of
zero Marangoni number $\left|\frac{\partial \gamma}{\partial T}\right| \frac{a^2 A
\rho_1}{{\eta_1}^2}$ (i.e., with negligible convective transport effects) the
droplet velocity ${\bf U}$ in the laboratory coordinate frame  is related to the
uniform temperature gradient by
\be
{\bf U} = \frac{a\: {\bf A}}{\eta_1 }\left|
\frac{\partial \gamma}{\partial T}\right| \bar U, \; \; \; 
\bar U = \frac{2}{(2+3\:\eta)\;(2+\kappa)}.
\label{eq1}
\ee
Here $\eta=\eta_2/\eta_1$ and $\kappa=\kappa_2/\kappa_1$ are the ratios of 
viscosities and thermal conductivities between the internal and the surrounding
fluids, respectively, $\left|\frac{\partial \gamma}{\partial T}\right|$ is a
constant derivative of the surface tension with respect to the local temperature
$T$ on the interface, and $\bar U$ is a dimensionless migration velocity. In the
case when the droplet and the external fluid have equal viscosity and thermal 
properties $\bar U=2/15 \simeq 0.133$. A review of the further development of the
subject  can be found in \cite{Sub92}.

Bratukhin \cite{Brat75} considered the migration of a deformable viscous 
droplet at the small Marangoni number. Using perturbation techniques in a
spherical coordinate system $(r, \theta)$ with origin at the center of
the  droplet, he found an expression for the  deformation of the droplet's shape
$r=R(\theta)$:
\begin{eqnarray}
R(\theta)& =  & 1 + Ma^2 s_2 P_2 (\theta) + O(Ma^3),  \nonumber\\
s_2&=&-\frac{3}{4 \gamma_0} \left( \frac{5}{2} {\bar U}^2 \frac{(1 - \rho)(1 +
\eta)}{4 + \eta} + \frac{Pr {\bar U}}{63 \chi}\frac{49 \chi-7\chi \kappa +
18\kappa}{(3 + 2\kappa)(2 + \kappa)} \right).
\label{eq2}
\end{eqnarray}
Here $Pr=\nu_1/\chi_1$ is the Prandtl number, $\rho=\rho_2/\rho_1$ and
$\chi=\chi_2/\chi_1$ are the ratios of density and thermal diffusivity
between the internal and the surrounding fluids, respectively, $\gamma_0 =
\gamma \rho_1 a/\eta_1$ is the dimensionless surface tension corresponding to the 
center of the drop, and $P_2 (\theta)$ is the second order Legendre polynomial. 

The first term of $s_2$ is related to  the inertia term in the Navier Stokes
equation, while the second one is related to the convective term in the energy
transport equation. It is interesting to note that the expansion of the
droplet deformation begins with the term of $O(Ma^2)$, and that
this term vanishes in the case that combines equal densities of the phases
($\rho=1$) and negligible convective heat transport ($Pr=0$). It follows from
[\ref{eq2}] that the deformations are small also for large values of the
reference surface tension $\gamma_0$ (small capillary number). 

Balasubramaniam and Chai \cite{BC87} showed
that the Young solution satisfies the full Navier-Stokes equations if the
temperature field on the spherical interface is proportional to $\cos \theta$,
i.e. in the case of negligible convective transport. The perturbation of the
spherical form constructed in \cite{BC87} under the assumption of small capillary
number has the same form as [\ref{eq2}].   

In real applications, however, fluid drops are not isolated and their motion is
influenced by the presence of neighboring particles. This interaction of drops
substantially alters the migration velocities and it may cause drops'
deformations {\it even in the case of negligible convective transport}. On the other
hand, the second mechanism of the deformations' suppression (large surface
tension) is expected to be valid in the multi body case as well.   

Most of the numerous theoretical studies of the interaction of bubbles and droplets
in the course of their thermocapillary migration were performed under the
assumption of nondeformable drops (zero capillary number). Meyyappan {\it et al} \cite{MWS} and
Feuillbois \cite{F} studied the Marangoni induced axisymmetric
migration of two bubbles  and Meyyappan and Subramanian
\cite{MS} studied the  non-axisymmetric case. Acrivos {\it et al} \cite{AJS}, 
Satrape \cite{Sat} and Wang {\it et al} \cite{WMA} examined ensembles of more than two bubbles.
The remarkable finding of these studies is that a bubble
migrates in the presence of another bubbles of the same size with exactly the same
velocity as when isolated. Bubbles of different sizes affect the motion of the
neighboring bubbles. The influence of the presence of a larger bubble on the motion
of a smaller one is more pronounced than its influence on the motion of the larger
bubble.  

These analyses were extended to the case of liquid drops by Anderson \cite{And85},
Keh and Chen \cite{KC1,Keh90,KC3} and Wei and Subramanian \cite{WS}. In 
this case, the migration velocity of a drop in the presence of another drops
differs from the velocity of an isolated drop even when the drops are of equal
size. Similar to the case of interacting bubbles, a larger drop exerts more
pronounced influence on the motion of a smaller one. The thermocapillary motion
of two non-conducting drops was studied by Loewenberg and Davis \cite{LD93}. 

The literature on the thermocapillary  motion of deformable drops is limited in
contrast to a somewhat related problem concerning the interaction of drops
undergoing a buoyancy driven motion. For the latter case numerous recent
studies revealed a rich variety of interaction patterns of deformable drops
depending on the Bond number and the initial configuration of the system ( see
e.g. Manga and Stone \cite{MS93}, Zinchenko { \it et al} \cite{Zin99}) and the
literature cited). Numerical simulations of an axisymmetric buoyancy-driven
interaction of a leading drop and a smaller trailing drop were recently
reported by Davis \cite{Dav99}. It was demonstrated that the trailing drop
elongates considerably due to the hydrodynamic influence of the leading
one. Afterwards, depending on the governing parameters,  the drops may either
separate and return to spherical shape or  the trailing drop may be captured by the
leading one, or one of the drops may break up. 

For the thermocapillary induced  motion, the effect of deformability was studied
mostly by a perturbation technique assuming small deformations. Recently,
Rother and Davis \cite{RD99} applied lubrication approximation to study the effect
of a slight deformability of the interfaces on the thermocapillary-driven migration
of two drops at close proximity.   

Zhou and Davis \cite{ZhD} presented a recent study in which the drops are free to deform 
in the course of their thermocapillary interaction. They  used a
boundary-integral technique to study the thermocapillary interaction of a
deformable viscous drop with a larger trailing drop making no {\it a priori}
assumptions  regarding the magnitude of deformations. It was demonstrated that in
this case the deformations are small and have a small effect on the drops motion at
moderate separation. The influence of deformability becomes significant only when
the drops are close together. Zhou and Davis \cite{ZhD} did not adjust the surface
tension to follow  the migration of the drops in the external temperature field
that would be stationary in the absence of a drop. Hence, the boundary conditions
that they used were not compatible with the continuous change of the positions of
the drops. 

The goal of the present work is to extend the analysis of the thermocapillary
interaction of deformable drops  to the cases of  equal-sized drops and the leading
larger drop and to  study the  influence of  the deformations on the relative
motion of the droplets. We have applied the boundary integral method for the
simulations of the thermocapillary induced motion of two deformable viscous
drops in the case of moderate capillary numbers and equal viscosity and thermal
properties of the dispersed and continuous phases.

\section{Problem formulation}

Consider two drops embedded in an immiscible viscous fluid with a temperature
gradient applied along their line of center.  In the absence of gravity
the drops migrate in the direction of the applied  temperature gradient one beyond
another due to thermocapillary forces. The trailing and the leading drop and the
continuous fluid are marked by indices $1$, $2$ and $3$, respectively. All the
fluids are assumed to be Newtonian and incompressible.  The interface tension
$\gamma$ is assumed to depend linearly on temperature
$\gamma(T) =\gamma(T_0)-\frac{\partial \gamma}{\partial T}(T-T_0)$. The
temperature field approaches a given linear function $T=Az+T_0$ far from the
dispersed species.  The problem is considered in a laboratory coordinate system, 
see Fig. 1.

Non-dimensional variables are introduced using the following scaling:  the radius
$a_1$ of the first droplet for the length, $V^*=\left|\frac{\partial
\gamma}{\partial T}\right|a_1 A/\eta_3$ for the velocity, $a_1/V^*$ for the time,
$\left|\frac{\partial \gamma}{\partial T}\right|A$ for the pressure and 
$Aa_1$ for the temperature. The problem is governed by the following dimensionless
parameters: the  Marangoni number, $Ma$, the Prandtl number,  $Pr$, and the
capillary number, $Ca$, where

\be
Ma=\left|\frac{\partial \gamma}{\partial T}\right| \frac{{a_1}^2
A\:\rho}{{\eta_3}^2};\;\;\; Pr=\frac{\nu_3}{\chi_3};\;\;\;Ca=\left|\frac{\partial
\gamma}{\partial T}\right| \frac{a_1A}{\gamma_0}.
\ee
The other dimensionless parameters are the ratios of the material properties of the
fluids, $\eta= \eta_i/\eta_3, \; \; \rho_i/\rho_3, \; \; \kappa=\kappa_i/\kappa_3$ and the
geometric parameters $R=a_2/a_1$ and $d={\cal L}/a_1$, where ${\cal L}$ is the
distance between the drops along their line of centers. 

We shall be interested in the case of small Marangoni numbers and moderate values
of the other parameters.  Following \cite{Brat75} one can see that, at the leading
order in $Ma$, the inertial and convective transport effects are negligible, and
the motion inside and outside the droplets is governed by the Stokes equations:
\begin{eqnarray}
\nabla p_3 &=& \triangle {\bf v}_3, \;\;\;{\bf x} \in \Om_3, \nonumber\\
\nabla p_i &=& \eta \triangle {\bf v}_i,\;\;{\bf x} \in \Om_i, \; \; i =1,2,
\nonumber\\
\nabla {\bf v}_i &=& 0, \;\;\;{\bf x} \in \Om_i, \; \; i =1,2,3,
\end{eqnarray}
where $p_i$ and ${\bf v}_i$ are the pressure and velocity field, respectively.
The temperature field inside and outside of droplets satisfies the
Laplace equation
\begin{equation}
\triangle { T}_i = 0, \; \;  \;   {\bf x} \in \Om_i, \; \; i =1,2,3.
\end{equation}
The continuous fluid is quiescent far from the drops
\be
{\bf v}_3  \to  0, \;\; \mbox {as}\;\; x\to\infty. 
\ee
At the interfaces the velocity field is continuous,
\be 
{\bf v}_{0}{\bf(x)} = {\bf v}_{i}{\bf(x)},\;\;{\mbox {for}} \;\;{\bf x
}\in\Sigma_{i},\;\; i=1,2,
\ee
and the kinematic boundary conditions require that
\be
\;\;\;\;\;\; \:{\bf n}_i\cdot{\bf v}_i = {\bf V}_{ni}
,\;\;{\mbox {for}}\;\;{\bf x }\in \Sigma_{i},\;\; i=1,2,
\label{kin}
\ee
where $V_{ni}$ is a velocity of the interface $\Sigma_{i}$ in the normal direction.
The tangential stress balance may  be written as
\be
\mbox{\bm $\tau$}_i {\bf \cdot} (\mbox{\bm $\Pi$}_{\rm 3} - \mbox{\bm $\Pi$}_{
i}) {\bf \cdot n}_i = - \frac{\ds{\pa T_i ({\bf x})}}{\ds{\pa \tau}},  \; \;
\; {\bf x} \in \Sigma_i, \; \; \; i =1,2, 
\ee
and the balance of the normal stresses reads
\be
{\bf n}_i \cdot (\mbox{\bm$\Pi$}_{\rm 3} - \mbox{\bm$\Pi$}_i) {\bf \cdot n}_i  =
\left( \frac{1}{Ca}- T_i({\bf x}) \right) \: H_i,  \; \; \; {\bf x} \in \Sigma_i,
\; \; \; i =1,2,
\ee
Here ${\bf \Pi}_3 = -p_3 {\bf I}+{\bf T}_3 $ and ${\bf \Pi}_i = -p_i {\bf I}+\eta \: {\bf T}_i $ with ${\bf T}_i $
being  the rate of deformation tensor,  $\mbox{\bm $\tau$}_i$ is a unit vector
tangential to the $i^{th}$ interface,  and $H_i=\nabla\cdot {\bf
n}_i$ is  the mean curvature on the  $i^{th}$ interface.

The temperature and the heat flux are continuous 
\be
T_i=T_3, \; \; \; \kappa\frac{\partial T_i}{\partial n}
= \frac{\partial T_3}{\partial n} \; \; \;
\mbox{for}   \; \; \; {\bf x} \in \Sigma_i, \; \; \; i =1,2,
\ee

and at infinity the temperature tends to a given linear function
\be
T \to z \; \; \; \mbox{at} \; \; \; {\bf| x|} \to \infty. 
\ee
 
\section{Boundary-integral formulation. Numerical method}

The boundary-value problem outlined in the previous section may also be formulated
in an equivalent integral form \cite{Poz92}. Following the classical potential
theory \cite{Lad} we can obtain a system of boundary-integral equations for the
temperature and velocities on the interfaces. When the thermal properties and
viscosities of the phases are equal for all $\Omega_i$ phases the boundary integral
equations simplify considerably, and the solution to the problem acquires an
explicit form. Thus, the temperature and the velocities everywhere are given by 
\be
T({\bf x}) = z,
\ee
\ba
{\bf v}({\bf x}) &=&
- \frac{1}{ 8 \pi} \oint_{\Sigma} {\bf J \cdot} \left[ \left(
\frac{1}{Ca}-T({\bf y}) \right) H({\bf y}) {\bf  n} +\frac{\ds{\pa T ({\bf
y})}}{\ds{\pa \tau}} \mbox{\bm  $\tau $} \right] d S({\bf y}). \label{biv}
\ea
Here ${\bf y} \in \Sigma=\Sigma_1+\Sigma_2$ and the kernels for the single  layer
potentials for Stokes flow are
\[
{\bf J}(\mbox{\bm $\xi$})= \frac{ \bf I}{\xi} + \frac{\mbox{\bm $\xi
\xi$}}{\xi^3}, 
\]
with $\mbox{\bm $\xi$}={\bf x} - {\bf y}$ and $\xi=|\mbox{\bm $\xi$}|$.
Note that the dimensionless surface tension  equals  $1/Ca-z$, and the natural
restriction $\gamma \ge 0$ implies that unequality $z \le 1/Ca$  hold for all
points on the interfaces during the entire time of the process. Applying this
unequality to the initial moment $t=0$ provides an upper bound for the capillary
number
\be
Ca < \frac{1}{2 R_1+2 R_2+ d_0}, \label{Cab} \ee 
where  $d_0$ is an initial separation between the surfaces of the drops.    

In the axisymmetric case, the integrations over the azimuthal angle in [\ref{biv}] 
can be performed analytically, and the problem is reduced to the computation of 
a one-dimensional singular boundary-integral for the interfacial velocity

\be
{\bf v}({\bf x})= \int_{\sigma} {\bf B}({\bf x},{\bf y})  {\bf \cdot f}({\bf
y})dl({\bf y}). 
\label{bivv}
\ee
 Here ${\bf x}=(r_x,z_x)$ and ${\bf y}=(r,z)$ are vectors on an azimuthal $(r,z)$
plane in a cylindrical coordinate system, $\sigma = \sigma_1+\sigma_2$ is the union
of curves describing the drops' surfaces in this plane, and $dl({\bf y})$ denotes
the differential arc length of the corresponding contour.

\[{\bf f}({\bf y})=\left( \frac{1}{Ca}-T({\bf y}) \right) H {\bf  n} +\frac{\ds{\pa
T ({\bf y})}}{\ds{\pa \tau}} \mbox{\bm  $\tau $} =f_n {\bf n}+f_\tau \mbox{\bm
$\tau $} \]
is a traction jump across the interface. The complete expressions for components
of the kernel $B({\bf x},{\bf y})$ are given in the Appendix. Note that the kernel
[\ref{bivv}] have a logarithmic singularity at ${\bf y} \to {\bf x}$.  In the
absence of Marangoni effect, when the tangential stresses are continuous ( $\pa T/
\pa \tau=0$), these singularities are often eliminated making use of a subtraction
technique \cite{Poz92}, that is based on a well known equality  
\be
\int_{\sigma_i} {\bf B}({\bf x},{\bf y})\cdot {\bf n}({\bf y}) d l ({\bf y}) =0.
\label{Beq}
\ee
When the tangential Marangoni stresses associated with the temperature
gradients on the drops surfaces are present, this procedure cannot be simply
incorporated into the singularity-subtraction formulation. For the purpose of  a
study  of the droplets migration it is sufficient to compute the normal
component of surface velocity $V_n={\bf n}(\bf x)\cdot {\bf v}({\bf x})$. We
employed a subtraction technique \cite{Poz92,LH96} in which the  normal component
of the velocity on the interfaces can be evaluated as

\[ V_n={\bf n}({\bf x})\cdot {\bf v}({\bf x})= {\bf n}({\bf x}){\bf
\cdot}\int_{\sigma} {\bf B}({\bf x},{\bf y}) {\bf \cdot f}({\bf y })dl({\bf y}) \]

\[
 = \int_{\sigma}f_\tau({\bf y }) \; {\bf n}({\bf x}){\bf \cdot  B}({\bf x},{\bf 
y}) {\bf \cdot} \mbox{\bm  $\tau $}({\bf y}) dl({\bf y}) \] 
\be + {\bf n}({\bf x}){\bf \cdot} \int_{\sigma} \left[ f_n({\bf y}) -f_n({\bf x^*} 
\right] {\bf B}({\bf x},{\bf y}) {\bf \cdot n}({\bf y }) dl({\bf y}), \label{vnt} 
\ee 
where ${\bf x^*}$ is an arbitrary chosen point in the bulk or on the
interface. The choice ${\bf x^*}={\bf x}$ removes a singularity in the
integral. The integrand of the first summand in the right-hand side of
[\ref{vnt}] tends to a finite limit at ${\bf y \to x}$ (see the Appendix). 

When the separation between the drops surfaces is small, 'near-singularities'
appear in the integrand if the reference point ${\bf x}$ is located in the gap
region. Following Loewenberg and Hinch \cite{LH96} we facilitate the calculation of
integrals at such 'near singular' using [\ref{vnt}], where ${\bf x^*}$ is the image
point across the gap region to ${\bf x}$.  This approach  provides a
regularization of the 'near-singularity' for the regions on the surfaces that are
in close proximity.

In our numerical computations, the drop boundaries in the azimuthal plane are
discretized using marker points and are approximated by cubic splines. The number
of points on the two interfaces is equal. The marker points are distributed
uniformly with respect to the arc length (non-uniform distribution was not needed
in our simulation because the curvature of the interface remains moderate at all 
time). The normal and tangent unit vectors at the interfaces and the mean
curvatures are computed by taking the derivatives of the position vector with
respect to the arc length. When the normal velocities on the interfaces are
obtained, the positions of the marker points are advances using Euler or Runge
Kutta method. After that the cubic spline approximation of the interfaces is
reconstructed and the marker points are redistributed on the new surfaces uniformly
with respect to the arc length.  The Krasny \cite{Krasny86} filtering procedure 
was used to smooth the spatial frequency spectrum and to remove noise that 
was introduced  by the computation of high derivatives in the code. All Fourier
modes below some 'tolerant number', which depends on the number of marker points
were set to $0$. At every step the positions of the drops' centers of mass were
obtained. The velocities of the centers of mass were consequently calculated  via
numerical differentiation with respect to time. The calculations are stopped when
the interfaces of the drops intersect or when the surface tension at any point on
the leading drop becomes negative. 

\bigskip

\section{Results and discussion}

We consider the axisymmetric drift of two deformable droplets with different
size. For simplicity we assume that  the droplets and the continuous media  have 
identical viscosity, densities and thermal properties. This ensures that the
temperature field is continuous and has continuous derivatives across the
interfaces. The droplets are initially spherical and start to move from an initial
separation distance. The parameters considered include the capillary number $Ca$, the
initial droplets' radii $R_1$ and $R_2$  and the initial distance $d_0$ between
surfaces of the droplets. During the motion we observe the evolution of the
positions and of the form of the droplets, and the flow patterns inside the
droplets and in the external fluid. The relative  position was measured  by the
distance  between centers of mass $L(t)$ and by the distance between the nearest
points on the surfaces of the two drops $d(t)$.  The deformation of a droplet was
characterized by the Taylor deformation parameter   
\be 
D=\frac{2 r_{max} - \Delta z}{2 r_{max} + \Delta z}, \label{tdf} 
\ee
where $\Delta z$ is the length of the drop along the axis of symmetry, and
$r_{max}$ is the maximum radial dimension. Positive and negative values of $D$
correspond to oblate and prolate shapes, respectively.

\bigskip

\subsection{Migration velocities. The case of non-deformable drops}

We have tested our numerical code by comparing calculated drift velocities with 
available analytic results. As it was mentioned above, there are analytical
results \cite{And85,Keh90} for the thermocapillary induced motion of non-deformable
spherical drops in a uniform external temperature gradient. These results can be
applied for the axisymmetric cases of drift of drops with different radii $R_1$ 
and $R_2$, and for different separation distances between the centers of mass. 
 
For the test of our code we considered a case of small capillary number,
$Ca=0.001$, in which the expected deformation of initial spherical shape
of the drops during the evolution of the positions is negligibly small. The case
of well-separated non-deformable drops was consider by Anderson \cite{And85}. He
showed that for droplets of equal radii, the velocities are equal  and have the
form

\begin{equation}
U = \left( 1 + \frac{1}{L^2} - \frac{7}{3} \cdot \frac{1}{L^6 }
\right) |{\bf U}|.
\label{and}
\end{equation}
 Keh and Chen \cite{Keh90} considered also drops at small distance and derived
a general formula for the velocities
\begin{eqnarray}
U_1 &=& M_{11}|{\bf U}_1| + M_{12}|{\bf U}_2|  \nonumber\\
U_2 &=& M_{21}|{\bf U}_1| + M_{22}|{\bf U}_2|
\label{keh}
\end{eqnarray}
where $M_{11},M_{12},M_{21},M_{22}$ are mobility coefficients that were
calculated numerically, and $|{\bf U}_1|,|{\bf U}_2|$ are droplets'
velocities that each drop would have in the absence of another drop.
 
The velocities of equal droplets calculated using the formulae [\ref{and}], 
[\ref{keh}] and our Boundary-Integral code are depicted in Fig. 2  versus the
separation distance. The good agreement in all distance-limits is visible. In the
case of large separations our calculations, denoted by solid circles, coincide with
Anderson's asymptotic formula (curve d), and in the limit of infinite distance
between the drops we recover the value of the drift velocity $0.133$ as in the
Young-Bratukhin problem \cite{YGB59,Brat75}. In the limit of a small separation
distance our results are in good agreement  with the calculations of Keh and Chen
as well (dashed curve). A comparison for the case of drops of different radii is
presented in Fig. 3.  We observe that the case of the drift velocities  of the two 
non-deformable different droplets is in a good agreement with Keh and Chen
predictions.  

\subsection{Deformation of equal-sized drops. Effects of Ca and velocity patterns}

Consider now the case of larger capillary numbers, where considerable deformations
are expected. The time evolution of the Taylor deformation parameter is shown in
Fig. 4 for different values of the capillary number and of the initial separation
distance. We observe two qualitatively different scenarios of the deformation
development.  For sufficiently large initial separations and small $Ca$
(e.g. curves d and e), the deformation is small and becomes almost stationary
after a short transient  period. On the other hand, for the relatively small
initial separations and large $Ca$ (e.g. curves a and b), the evolution of the
deformation after the transient period becomes slower, but remains essentially
non-steady. During the initial period, with time scale of $O(Ca)$, the deformations
occur because   the non-uniform surface tension is not compatible with the initial
spherical form of the drop.  The motion during this period is driven by the surface
tension, rather than by its gradient, and the deformation is controlled by the
normal component of the interfacial stress. At longer time scales the development
of the deformations is induced mostly by the changes in the mutual positions of the
drops  (characterized by the separation distance). In Fig. 5  the deformation
parameter is depicted versus separation distance for equal-size drops and
different capillary numbers. The  initial separation was kept constant at
$L_0=2.05$. The leading drop becomes more deformed than the trailing one. This can
be explained by the fact that the leading drop is located in a hotter region and,
thus, it has a lower surface tension.  The dependence of well developed
deformations on the capillary number is demonstrated in Fig. 6. Here we plotted the
deformation parameter of the more deformed leading drop in Fig. 5 when the
separation distance reached $L=2.04$. The dependence on small $Ca$ appears to be
linear.

The deviation of the surfaces from the spherical shape $\sqrt{r_i^2(\theta)+
(z_i(\theta)-Z_i)^2}-R_i$ is demonstrated in Fig. 7 for different values of
time. One can see considerable deformations in the gap region while the outer
regions of the drops remain almost spherical. It is evident from Figs. 4, 5
and 7 that the leading drop deforms to an oblate shape while the trailing one 
becomes prolate. The mechanism of this deformation is shown in Figure 8 where we
plot the velocity field for two equal drops at close proximity 
for time $t=10$. The reference frame moves with the rear  point on the axis of the 
trailing drop as marked in the figure. In this reference frame, the drops
are almost at rest except for the gap region. The motion is composed of a flow
past an aggregate of two drops and the deformation of the gap region in the
opposite direction. The elongation of the trailing drop to a prolate shape
and the flattening of the leading drop to the oblate shape are clearly evident.

Similar to the deformation, the velocities of the drops' centers of mass achieve
quasi-steady slowly changing values after a short transient period. For two
equal drops at $Ca=0.2$ these are plotted in Fig. 2 versus the distance between the
surfaces (points on the curves $a$ and $c$).  One can see that, at large separations,
the velocities of the two equal-sized drops are almost equal and approach the
velocities of the non-deformable drops.  In contrast to this, for sufficiently
small separation, the velocities of the two deformed drops are different. The
leading oblate drop moves slower than the trailing prolate one.  

The time evolution of the surface-to-surface distance $d(t)$ and center-to
center distance $L(t)$ for $Ca=0.2$ and different initial separations is shown in
Fig. 9. We observe two different behaviors depending on the initial separation. At
small initial separations, surface to surface distance decreases with time (Figure
9 (e)). For larger separations, surface-to surface distance shows a transient
behavior in which it grows initially and decreases after some period
(Fig. 9(b)). This initial period of repulsion increases with the initial
separation, and for $d_0=0.5$ (see Fig 9(c)) the surface-to-surface distance grows
during the entire time of the process. To understand this behavior recall that the
leading drop deforms into an oblate shape, thus, moving apart from the trailing
one. The latter, in turn, deforms into a prolate shape, however, its deformations
are lower due to a higher surface tension. As a result, the separation distance
increases. The difference in the effective surface tension increases with the
separation distance, hence, when the distance is sufficiently small, this effect is
overcome by the simultaneous approach of the droplets' centers of mass (Fig. 9 (d)
and (f)) and, after a period of repulsion, the drops surfaces approach each
other. The period of initial 'repulsion' becomes wider for larger initial
separations, while when the initial distances are small enough,  the difference in
the rate of deformation becomes negligible, the initial repulsion period vanishes 
(see Fig. 9 e) and the drops approach during the entire time of the process. Note
that the motion of the droplets towards each other is quite slow, and in the cases
that we considered  drops that were initially very close together do not progress
to a complete collision.

\subsection{Interaction of Drops of Different Size.} 

The interaction of equal size drops is governed by the interplay of two  
effects: relative motion caused by the oblate(prolate) shape of the 
leading(trailing) drop, and more pronounced deformations of the leading drop in
the gap region due to its lower surface tension. These effects take place also for
unequal drops, but in this case they are combined with the relative motion caused
by the different sizes.  Similar to the case of equal drops, after a short
transient period the migration velocity of the trailing drop is enchanced and the
leading drop motion is retarded compared to the case with no deformations. If the
leading drop is smaller than the trailing one, the separation distance decreases
with time and the drops come to near contact. For this case, the larger drop
remains almost spherical, and the smaller one deforms considerably.  The form of
the drops in close proximity and the velocity field is shown in Fig. 10 (a) for
$Ca=0.2, \; \; \; R_1=1,  \; \; R_2=0.5$. The reference frame moves with the rear
point of the trailing drop.  The motion is a superposition of a flow past an
aggregate of the drops and a motion and deformation of the surface of the leading
drop. The larger trailing drop deforms mostly in the near gap region. The
deformation pattern in this case agrees qualitatively with the one reported by Zhou
and Davis \cite{ZhD}.

The situation is different when the trailing drop is smaller than the leading
one. In this case the gap between nondeformable drops increases and the 
drops separate. The deformability may result in a more complex interaction between
such drops if the initial separation is small enough.   The velocity field is shown
in Fig. 10 (b) for $Ca=0.2, \; \; \; R_1=1, \; \; R_2=1.5$. The reference frame
moves  with the point at the center of the gap region. One can observe a very slow
motion in the gap and considerable normal velocities at the interfaces in the outer
regions, indicating the developing deformations. The evolution of the separation
distances  and the deformation parameter are shown   in Figs 11, 12 and 13    
for various drop size ratio and $Ca$. Different curves on each plot correspond
to different initial separations.  It is seen that for the large initial
separations, the surface-to-surface distance increases during the entire
duration of the process, as in the case when the deformations are
absent.  When the initial separation is small the distance is first reduced
and only increase at the later stages. The transition is controled by the capillary 
number (see Fig.14) and it seems to occur at separation
distance of $O(10^{-2})$. 

To understand this unusual behavior, consider the evolution of the deformation
factor that is shown in these figures.  The smaller drop deforms much
stronger than the larger one assuming a prolate form. This is an anticipated result
since the deformation of the trailing drop is induced by the pulling presence of
the leading one and vice versa.  The larger drop induces larger deformations. Since
in our case the leading drop is larger, higher deformations of the trailing drop
are expected. At the depicted size ratios this effect overcomes the effect of
a lower surface tension over the leading drop that has an opposite effect in the
case of equal drops, where the leading drop is more deformed.  

The deformation of the smaller drop develops very fast during an initial transient
period and becomes slower afterwards. The interface of a larger drop is initially 
pushed by the trailing one and assumes an oblate shape.  These deformations  are
combined with the relative motion of the drops apart from each other due to the
different radii. Unless the drops are very close in size, after the initial period
the elongation of the trailing drop becomes slower, and the distance between the
interfaces begins to grow. The drops separate and the influence of the trailing
drop on the leading one causes its deformation to reverse itself and it also
becomes prolate (see Figs. 12(c)  and 13(c)). When the drops are very close in size
(see  Fig. 11(c)), the evolution their shape is more symmetric and resembles the
one  depicted  in Fig. 4 for the equal-size case. 

During the initial period the drops may come very close together and attractive
London van der Waals forces may results in the rupture of the inter-droplet film and
cause an eventual coalescence. This process, however,  is beyond the scope of our
work. The distance between the centers of mass of the drops increases monotonically
while the surface exhibits a transient behavior. During the
early stages of the dynamics the gap shrinks and the surfaces come to close
proximity, while at later stages the drops separate as in the case with no
deformations. The time in which the initial perturbation of the shape is relaxed
and the gap acquires a minimum appears to depend linearly  on the capillary
number. In this case the system may result in a higher coalescence rate due to
diminishing gap in the transient period.

\section{Concluding remarks}

The interaction between two drops in a temperature gradient shows a variety of
behavior patterns. Equal drops always  approach each other, however, the  
surface-to-surface distance has a transient behavior during which the separation
first grows and later diminishes. When two unequal drops are moving, with the
bigger one leading, an opposite transient behavior is observed if the initial
separation distance is small. Here, the surfaces first approach each other and
later depart while the drops are moving away from each other. This non-trivial
behavior takes place during an initial period of the process. Since our
calculations are based on the creeping flow approximation that neglect inertia of
the fluid, these results are applicable only if the flow field in the fluid is
established much faster than  an equilibrium form of the deformed surface. Three  
different time scales are present in the process  under consideration: a viscous
relaxation time  $t_{\nu} = a_1^2\rho/\eta_3$, capillary surface relaxation time
$t_{s}=\eta_3 a_1/\gamma_0$,  and the thermocapillary time $t^*= \eta_3
/\left(\frac{\partial \gamma}{\partial T} A \right)$, that was chosen as the
time scale in our considerations.   Since $ t_{\nu}/ t^*= Ma$ and $t_s/t^*= Ca$,
the condition $Ma << Ca$ is sufficient to ensure that the transient period
can be described in a creeping flow approximation. 

It is interesting to compare the magnitude of deformation in our case with
different cases such as buoyancy-driven motion \cite{Dav99} and that resulting from
non-linear dependence of surface tension on the external field \cite{Adams76}. We
found that the deformation in the case of a thermocapillary migration causes
behavior of the drops at close proximity that is qualitatively different from that
observed in the other cases.  The calculated magnitude  in the thermocapillary was
relatively small for both leading and trailing drops and for different aspect
ratios. The rich variety of interaction pattern typical for a motion driven buoyancy
is not observed. There are several reasons why it is so. The first is a well-known 
fact that the disturbances of the velocity field, resulting from the
thermocapillary-induced motion of a single drop, decay with the distance from the
particle as $1/r^3$. This is  much faster  than the $1/r$ decay in the case of the gravity-driven
motion. Thus, the influence of the presence of a drop on the motion and
deformation of the other one is expected to be smaller and to decay more rapidly
with the separation distance than the one in the gravity-driven motion case. 
The other important difference of these two processes is that the Bond number
 \[
Bo=\frac{\Delta \rho g {a_1}^2}{\gamma_0},
\]
that governs the deformability of the drops undergoing gravity-driven migration,
may take any values from $0$ to infinity, and the interesting behavior is
observed when this value is relatively large, $Bo>3$ (see \cite{Dav99}).   In
contrast to this, the value of the capillary number that governs the deformability
of the drops in the course of their  thermocapillary-driven migration is
restricted from above if only physically relevant cases of positive interfacial
tension are considered. In our calculations the non-vanishing values of $Ca$ were
typically about $0.2$. 
 
Note that using the linear profile of a surface tension -- temperature dependence 
 results in restriction on the extent of collective drift of
drops. Practically there is some moment of time at which the calculation must 
stop, because after this time the surface tension becomes zero at the front
edge of  the leading drop. In the case  of temperature dependence  this
situation  implies a phase transition. However, when the surface tension
depends on concentration of a solute, zero surface tension  cannot be
achieved.  In this case the dependence of surface tension on concentration (
for some pairs fluid-solute) can be divided to two different regions. In the
first region the dependence of tension on concentration is strong and can be
described by the linear model considered above.  In the second region, when
the surface tension is diminished,  its change with concentration is very slow 
and it can be approximated  by a constant \cite{Adams76}.

We performed  a calculation for the case of equal drops using this simple two-region
dependence of the interfacial tension letting the linear dependence  down the value
$\gamma / \left( A a_1 \frac{ d \gamma}{d T} \right)=10^{-2}$ and keeping it
constant beyond this point.  When the drops enter the region of low surface tension
the deformation pattern
changes and more pronounced distortion of the spherical shapes are obtained. This
deformation  pattern is accompanied by a clear reduction of the drops' speed and
the entire migration is halted. A demonstration of such behavior is given in Figure 15.
More detailed dynamics of such interactions are left for the future study. 

 \bigskip
 
\section*{{Acknowledgements}}
 
\medskip
 
This research was supported by The Israel Science Foundation founded by the
Academy of Science and Humanities. O. M. L. acknowledges the support of the Israel
Ministry for Immigrant Absorption. The authors wish to thank A. M. Leshansky for
helpful discussions.
 
 \bigskip

\section*{ Appendix: Kernels in Eqn. [16]}
 
\medskip
 
If $r_x \neq 0$, i.e. the point ${\bf x}$ is not on the axis of symmetry
 
\[
B_{zz}=2k \left( \frac{r}{r_x} \right)^{1/2} \left[
F(k)+\frac{(z-z_x)^2}{(r_x+r)^2+(z_x-z)^2} E(k) \right],
\]
 
\[
B_{zr}=k \frac{z-z_x}{(r_x r)^{1/2}}  \left[
F(k)-\frac{r_x^2-r^2 +(z-z_x)^2}{(r_x+r)^2+(z_x-z)^2} E(k) \right],
\]
 
\[
B_{rz}=-k \frac{z-z_x}{r_x}\left( \frac{r}{r_x} \right)^{1/2}
\left[ F(k)+\frac{r_x^2-r^2-(z-z_x)^2}{(r_x+r)^2+(z_x-z)^2} E(k)
\right], \]
 
\[
B_{rr}=\frac{k}{r_x r} \left( \frac{r}{r_x} \right)^{1/2} \left\{
[r_x^2+r^2+2(z_x-z)^2] F(k) - \right. \]
\[\left. \frac{2(z-z_x)^4+3(z-z_x)^2
(r_x^2+r^2)+(r_x^2-r^2)^2}{(r_x+r)^2+(z_x-z)^2} E(k) \right\},
\]
 
Here $F$ and $E$ are complete elliptic functions of the first and second kind
respectively,
 
\[k^2=\frac{4r_x r}{(r_x+r)^2+(z_x-z)^2}, \]
and $n_z$ and $n_r$ denote the projection of the normal vector in the azimuthal
plane on the axis $z$ and $r$.
 
If the point ${\bf x}$ is on the axis of symmetry, $r_x=0$, $k=0$ and the kernels
take the form
 
\[ B_{zz}=2 \pi r \frac{r^2+2(z-z_x)^2}{[r^2+(z-z_x)^2]^{3/2}}, \; \; \; B_{rz}=0,
\]
\[ B_{zr}=-2 \pi\frac{r^2 (z-z_x)}{[r^2+(z-z_x)^2]^{3/2}}, \; \; \; B_{rr}=0, \]

Asymptotic expansions as ${\bf y}$ tends to ${\bf x}$ ($r_x \neq 0$):
 
\[ B_{zz}=- \ln [r^2+(z-z_x)^2] +c_{zz}+ \ldots\,  \; \; \;
B_{rz}=c_{rz}+\ldots,
\]
 
\[ B_{zr}= c_{zr}+\ldots, \; \; \;
B_{rr}=- \ln [r^2+(z-z_x)^2] +c_{rr}+\ldots,
\]
 
\[ {\bf n}({\bf x}){\bf \cdot  B}({\bf x},{\bf
y}) {\bf \cdot} \mbox{\bm  $\tau $}({\bf y})=c_{rz} n_r({\bf x}) \tau_z({\bf x})
+c_{zr} n_z({\bf x}) \tau_r({\bf x})+\ldots, \]
 where $c_{rr}$, $c_{rz}$, $c_{zr}$, and $c_{zz}$ are constants.  If the curve
$\sigma_i$ is parametrized by the arc length, the kernels are regular for $r_x=0$.

\newpage
{\large \bf Figure legends}
\bigskip
\begin{enumerate}
\item[{\bf Figure 1.}] Geometric sketch of two viscous drops immersed in an
external temperature gradient $\nabla T$ parallel of their axis of symmetry.

\item[{\bf Figure 2.}] Center of mass velocities for equal drops $R_1=R_2=1$. The
solid line $(d)$, dashed line $(b)$ and the points on these lines  are the results
of Anderson, Keh and Chen and our computations for $Ca = 0.001$, respectively. The
points $(a)$ and $(c)$ correspond to leading and trailing deformable drops at
$Ca=0.2$. 

\item[{\bf Figure 3.}] Droplet velocities for $Ca=0.001$ and different aspect
ratio. In the case $(a,b)$ $ R_1= 1.0$ and $R_2=2.0$. In the case $(c,d)$ $ R_1=
1.0$ and $R_2=0.5$. The lines $b$ and $d$ correspond to the larger drop.   

\item[{\bf Figure 4.}] The evolution of Taylor deformation parameter defined in
Eqn. [\ref{tdf}]: $(a)$ $ Ca=0.2, \; d_0=0.1$; $(b)$ $ Ca=0.2,
\; d_0=0.5$; $(c)$ $ Ca=0.1, \; d_0=0.1$; $(d)$ $ Ca=0.1, \; d_0=0.5$; $(e)$
$ Ca=0.1, \; d_0=2$. Positive and negative values correspond to oblate and prolate
shapes, respectively. 

\item[{\bf Figure 5.}] Deformation parameter for equal-size droplets versus the
distance between centers of mass:  $(a)$ $ Ca=0.2, \; d_0=0.05$; $(b)$ $
Ca=0.1, \; d_0=0.05$; $(c)$ $ Ca=0.05, \; d_0=0.05$. Positive and negative values
correspond to oblate and prolate shapes, respectively.  

\item[{\bf Figure 6.}] Deformation of the leading droplet versus the capillary
number at $L=2.04$ for equal-size droplets.

\item[{ \bf Figure 7.}] The local deformations for trailing $(a)$ and leading $(b)$
equal-size droplets for the case $Ca=0.2$ and $d_0= 0.01$. 1, 2 and 3 correspond to
$0.5$, $3.3$, and 10 units of time, respectively.

\item[{\bf Figure 8.}] Velocity patterns for $Ca=0.2$, $R_1=1.0$ and $R_2=1.0$. The
reference frame moves with the speed of the marked point. 
 
\item[{\bf Figure 9.}] The evolution of the separation and center-to-senter
distances for equal-size drops at $Ca=0.2$ as function of the initial
separation $(a)$. $(b)$ Transient dynamics for intermediate initial separation. $(c,d)$
Evolution of surface-to-surface  separation and center-to-center distance for
relatively large initial separation. $(e,f)$ Evolution of surface-to-surface
separation and center-to-center distance for relatively small initial separation. 

\item[{\bf Figure 10.}] Velocity patterns for $(a)$ $Ca=0.2$, $R_1=1.0$ and
$R_2=0.5$;  $(b)$ $Ca=0.15$, $R_1=1.0$ and $R_2=1.5$. The reference frame moves
with the speed of the marked point. 
 
\item[{\bf Figure 11.}] Interaction between a leading larger drop and a trailing
smaller one at close proximity for $Ca= 0.2, \; R_1=1.0, \; R_2=1.1$: $(a)$
evolution of center-to-center distance; $(b)$ evolution of surface-to surface
separation; $(c)$ evolution of Taylor deformation parameter. The numbers 1 to 9
denote the different initial separation.

\item[{\bf Figure 12.}] Interaction between a leading larger drop and a trailing
smaller one at close proximity for $Ca= 0.15, \; R_1=1.0, \; R_2=1.5$: $(a)$
evolution of center-to-center distance; evolution of surface-to surface
separation; $(c)$ evolution of Taylor deformation parameter. The numbers 1 to 6
denote the different initial separation.
 
\item[{\bf Figure 13.}] Interaction between a leading larger drop and a trailing
smaller one at close proximity for $Ca= 0.12, \; R_1=1.0, \; R_2=2$: $(a)$
evolution of center-to-center distance; $(b)$ evolution of surface-to surface
separation; $(c)$ evolution of Taylor deformation parameter. The numbers 1 to 7
denote the different initial separation.
 
\item[{ \bf Figure 14.}] The relaxation time $tr$ with minimum separation distance
as function of the capillary number.  

\item[{ \bf Figure 15.}] Deformation of equal-size drops  migrating into a region
of vanishing surface tension. The upper plot shows the dependence of surface
tension on distance. The lower plot demonstrates migration and deformation
pattern for various time. $(a)$  $t=0$; $(b)$ 
$t=5$; $(c)$ $t=19$; $(d)$  $t=25$. 
\end{enumerate}

\end{document}